\documentclass[%
 reprint,
superscriptaddress,
bibnotes,
 amsmath,amssymb,
 aps, prb
]{revtex4-2}

\usepackage{graphicx}
\usepackage{dcolumn}
\usepackage{bm}
\usepackage{braket}
\usepackage[colorlinks=true, allcolors=blue]{hyperref}
\usepackage{cleveref}

\begin{document}
\preprint{APS/123-QED}

\title{“One defect, one potential” strategy for accurate machine learning prediction of defect phonons}
\author{Junjie Zhou}
\affiliation{College of Integrated Circuits and Micro-Nano Electronics, and Key Laboratory of Computational Physical Sciences (MOE), Fudan University, Shanghai 200433, China}
\author{Xinpeng Li}
\affiliation{College of Integrated Circuits and Micro-Nano Electronics, and Key Laboratory of Computational Physical Sciences (MOE), Fudan University, Shanghai 200433, China}
\author{Menglin Huang}
\email{menglinhuang@fudan.edu.cn}
\affiliation{College of Integrated Circuits and Micro-Nano Electronics, and Key Laboratory of Computational Physical Sciences (MOE), Fudan University, Shanghai 200433, China}
\author{Shiyou Chen}
\email{chensy@fudan.edu.cn}
\affiliation{College of Integrated Circuits and Micro-Nano Electronics, and Key Laboratory of Computational Physical Sciences (MOE), Fudan University, Shanghai 200433, China}

\begin{abstract}
Atomic vibrations play a critical role in phonon-assisted electron transitions at defects in solids. However, accurate phonon calculations in defect systems are often hindered by the high computational cost of large-supercell first-principles calculations. Recently, foundation models, such as universal machine learning interatomic potentials (MLIPs), emerge as a promising alternative for rapid phonon calculations, but the quantitatively low accuracy restricts its fundamental applicability for high-level defect phonon calculations, such as nonradiative carrier capture rates. In this paper, we propose a “one defect, one potential” strategy in which an MLIP is trained on a limited set of perturbed supercells. We demonstrate that this strategy yields phonons with accuracy comparable to density functional theory (DFT), regardless of the supercell size. The predicted accuracy of defect phonons is validated by phonon frequencies, Huang–Rhys factors, and phonon dispersions. Further calculations of photoluminescence (PL) spectra and nonradiative capture rates based on this defect-specific model also show good agreements with DFT results, meanwhile reducing the computational expenses by more than an order of magnitude. Our approach provides a practical pathway for studying defect phonons in 10$^4$-atom large supercell with high accuracy and efficiency.
\end{abstract}

\maketitle

\section{Introduction}

\par Phonon-assisted electronic transitions (multiphonon transition \cite{huang1981lattice}) at lattice defects play a crucial role in determining the performance of a wide range of semiconductor devices as well as quantum computing and communication systems \cite{huang1950theory,IgorSOsadko_1979,toyliEngineeringQuantumControl2013,awschalom2018,Gali2019}. In microelectronic and optoelectronic devices, carrier capture from the band edge into defect levels occurs via a multiphonon-assisted nonradiative process \cite{kubo1955application,passler1974description,passler1975description,huang1981lattice}, which directly affects carrier lifetimes and, in turn, impacts device operating speed, efficiency, power consumption, and reliability \cite{NonradiativeVoltage2017,GRASSER201239}. Accurate evaluation of phonon properties in defect systems, and thereby the rates of such nonradiative multiphonon processes, is therefore essential for device design. In defect color centers and qubit systems, electronic transitions between defect levels are accompanied by phonon emission in addition to photon emission, giving rise to sideband structures near the zero-phonon line in photoluminescence (PL) spectra \cite{RRSA1976}. These features serve as key fingerprints for the accurate identification of quantum defects. Consequently, precise calculations of phonon-induced PL sidebands are indispensable for quantum defect research. Taken together, predicting the nonradiative transition rates and the radiative PL spectra of defect systems both rely on accurate descriptions of defect-related phonons.
 
\par Although the theoretical framework of multiphonon transitions has been established \cite{huang1950theory,kubo1955application,passler1974description,passler1975description}, practical calculations are still computationally challenging. The main bottleneck lies in the huge computational cost of phonon calculations for defects using the supercell model and density functional theory (DFT). For a defect supercell containing $N$ atoms, phonon calculations using the finite-displacement method typically require $6N$ DFT self-consistent calculations in a brute-force manner, \textit{e.g.}, 1800 calculations are required for a 300-atom supercell. If phonons are needed in another defect charge state to account for phonon renormalization \cite{zhou2025}, this cost effectively doubles. These unavoidable computational expenses make it impractical to perform full-dimensional calculations of electron-phonon coupling in defect-induced multiphonon transitions for large supercells. Consequently, simplifications are often adopted, \textit{e.g.}, reducing the full set of $3N$ phonon modes to a single effective mode \cite{alkauskas2014first,turiansky2021nonrad}, under the assumption that this mode couples most to the lattice relaxation.

\par Recent advances in machine learning interatomic potentials (MLIPs) offer a promising alternative \cite{allegro2022,MACE_2022,yang2024mattersim}, demonstrating remarkable success in predicting defect energetics and, more recently, phonon frequencies \cite{Satoshi2022,zhangVacancy2024,Satoshi2025}. However, most MLIP efforts have focused on developing foundation models applicable to a wide range of properties and materials \cite{loewUniversalMachineLearning2025,LEE2025101688,du2025universalmachinelearninginteratomic,10.1063/5.0199743}. For example, Sharma et al. \cite{sharma2025} used universal MLIP models to compute PL spectra for 791 defects in 10 different 2D host crystals. The overall PL lineshape is reasonably well reproduced compared to DFT results, suggesting the potential of the foundation model in predicting phonon-related properties. 
However, they pointed out that the predicted Huang–Rhys factors for these defects deviate by about 12\%, and noticeable discrepancies remain in the detailed features of the PL spectra. This can be understood since the training of foundation models does not consider the local relaxation around defects. Thus, the phonon properties predicted from the foundation model, including phonon frequencies, eigenvectors, and Huang-Rhys factors still exhibit obvious errors. It is important to emphasize that even a small error in phonon frequencies and eigenvectors can be significantly amplified in the calculated PL lineshapes and nonradiative transition rates. Therefore, accurate prediction of defect phonon properties with low computational cost is still a bottleneck, limiting the study of multiphonon transitions \cite{wang2024perovsdopants,wang2025evaluating}.

\par In this paper, a “one defect, one potential" strategy is proposed to overcome this issue. We demonstrate that training a defect-specific MLIP offers an effective compromise between accuracy and computational efficiency for calculating phonon-related quantities. The local descriptor inherent in the equivariant model significantly enhances the training efficiency \cite{allegro2022,musaelian2023}, enabling reliable predictions with a relatively small training set, regardless of the supercell size. Using $\text{C}_\text{N}$ in GaN and $\text{Li}_\text{Zn}$ in ZnO as examples, we show that the phonon frequencies, eigenvectors, and Huang-Rhys factors predicted by the trained MLIP are in excellent agreement with those from DFT. These consistencies allow for an accurate modeling of phonon sidebands in PL spectra and nonradiative transition rates at the level of hybrid functional, while reducing the computational cost by orders of magnitude. This defect-specific strategy represents a paradigm shift from conventional approaches for calculating defect phonon properties in large supercells, and greatly enhances our ability to study more complex physical problems in multiphonon transitions at defects.

\section{Methodology}

\subsection{Neural network potential}

In order to achieve relatively accurate MLIP with a limited amount of training data, the Neural Equivariant Interatomic Potentials (NequLP) graph neural network framework is selected, which is based on E(3)-equivariant operators and is highly data-efficient in predicting interatomic potentials for molecules or materials \cite{Behler2007,allegro2022}.

In practical calculations, the \texttt{Allegro} package \cite{allegro2022} is utilised to construct a two-body latent multilayer perceptron (MLP) with hidden dimensions [64, 64, 128, 128, 128] and a later latent MLP with [128, 128, 128], both with SiLU nonlinearities. The two-body latent MLP cutoff radius is 6 Å, exhibiting full O(3) symmetry. The training set is comprised of 85\% of the total data set, with the remaining 15\% allocated for the validation set.

\begin{figure}[h]
	\centering
	\includegraphics[width=0.45\textwidth]{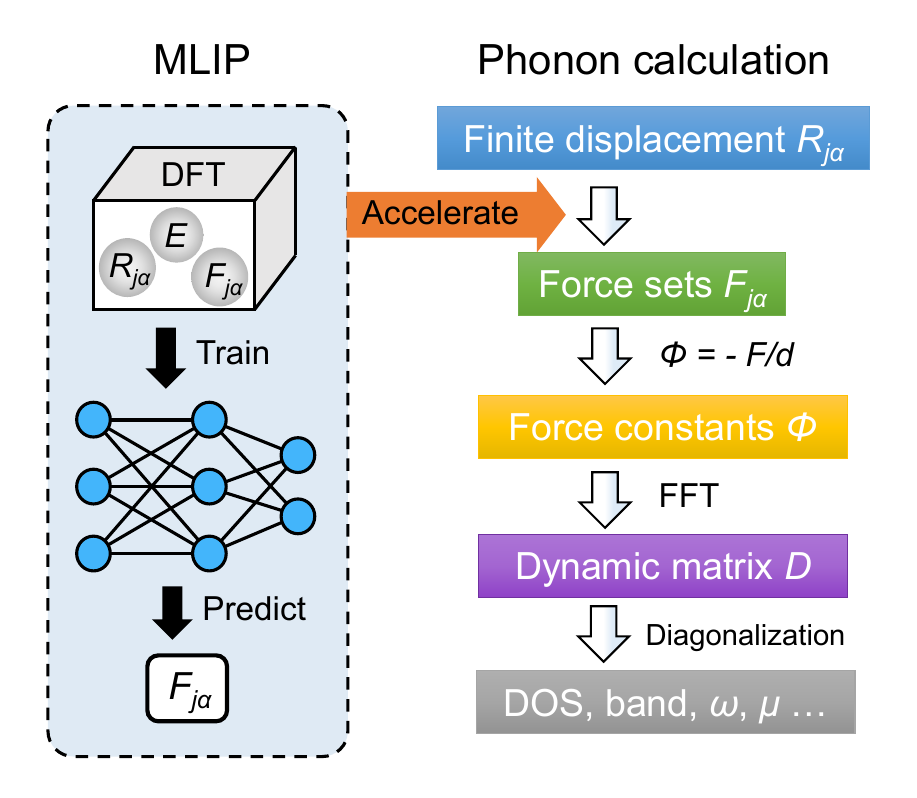}
	\caption{Workflow of accelerating phonon calculations through MLIP. The method is based on the finite displacement method, uses MLIP to predict forces.}
	\label{FIG1}
\end{figure}

\subsection{Training dataset generation}


The size of the training dataset plays a critical role in determining the accuracy of the MLIP. An insufficient amount of training data often leads to underfitting, resulting in errors that are incompatible with the precision required for phonon calculations. On the other hand, excessively large datasets yield diminishing returns in accuracy while significantly increasing computational cost, potentially eliminating the efficiency advantage over conventional DFT calculations. Therefore, achieving an optimal balance between training set size and accuracy is essential for the practical applicability of MLIP in phonon calculations.
We evaluated the effect of training set size and found that, for both the 96-atom and 360-atom supercells, as few as 40 sets (including those used for validation) were sufficient to achieve accurate prediction of phonon frequencies and eigenvectors. A more detailed analysis of the training set size and its impact on phonon accuracy is provided in the Supplemental Material.

For the first-principles calculations based on DFT, we employed the Vienna \textit{ab initio} simulation package (\texttt{VASP}) \cite{kresse1996efficiency,kresse1999ultrasoft} to obtain the reference data required for training the interatomic potential, including total energies $E$ and atomic forces $F$. Two types of defect were selected for training and validation: $\text{C}_\text{N}$ in GaN and $\text{Li}_\text{Zn}$ in ZnO. For both systems, the exchange–correlation functional was treated within the Perdew, Burke, and Ernzerhof (PBE) formulation \cite{PBE}, and the interaction between core and valence electrons was described using the projector augmented-wave (PAW) method \cite{PAW}. The plane-wave energy cutoff was set to 400 eV. To ensure convergence in the DFT-based phonon calculations, the force convergence criterion was set to 10 meV/Å for GaN and 1 meV/Å for ZnO during structural relaxation.

For simplicity, the training data were generated using supercells with the same size as those employed in the subsequent phonon calculations. This allows us to directly use the optimized structures without considering boundary forces arising from changed supercell. The generation of each training structure started from a relaxed structure, each atom in the supercell was randomly displaced within a sphere of radius $r_{\text{max}} = 0.04$ Å centered at its equilibrium position $\mathbf{R}_0$. Both the radial and angular components of the displacements $\Delta\mathbf{R}$ were sampled from uniform distributions. The choice of $r_{\text{max}}$ was informed by the displacement magnitude used in the finite displacement method for phonon calculations, which was set to 0.01 Å throughout this work. In the Supplemental Material, we present an analysis of the error in the force constants $\Phi$ as a function of the random displacement radius $r_{\text{max}}$, showing that a value of 0.04 Å provides a reasonable balance between perturbation magnitude and force accuracy.

\subsection{Phonon calculation}
The MLIP-accelerated scheme developed in this work can be seamlessly integrated into the conventional DFT-based phonon calculation workflow. The overall computational procedure is illustrated in Fig.~\ref{FIG1} and consists of two main stages.

In the first stage, the MLIP is trained. Using DFT, we calculate the total energy \(E\) and atomic forces \(F_{j\alpha}\) for defect-containing supercells \(R_{j\alpha}\) generated via random displacement, where the index \(j\) denotes the atom and \(\alpha = x, y, z\) represents the Cartesian coordinates. These quantities \(\{R_{j\alpha}, F_{j\alpha}, E\}\) serve as the training data for constructing the potential energy surface of the structure. Once trained, the MLIP is capable of predicting forces \(F_{j\alpha}'\) for any given atomic structure \(R_{j\alpha}'\).

The second stage involves phonon calculations. In this paper, \texttt{Phonopy} package \cite{phonopy-phono3py-JPCM,phonopy-phono3py-JPSJ} is applied to generate structures and implement phonon calculations. For the wurtzite structure like GaN and ZnO considered in this work, the finite displacement method requires approximately \(3N\) displaced supercells, each of which conventionally demands a separate self-consistent DFT calculation to obtain \(F_{j\alpha}\). This step typically dominates the total computational cost of phonon calculations. By contrast, the trained MLIP can predict the forces \(F_{j\alpha}\) for each displaced structure \(R_{j\alpha}\) within seconds, offering a significant speed advantage. Within the harmonic approximation, these forces are then used to construct the force constant matrix, which are computed using the finite displacement method as
\begin{equation}\label{force_constant}
    \Phi_{j\alpha,j^\prime\alpha^\prime} = -\frac{F_{j\alpha}}{d_{j^\prime\alpha^\prime}},
\end{equation}
where \(d_{j^\prime\alpha^\prime}\) denotes the displacement of atom \(j^\prime\) along the Cartesian direction \(\alpha^\prime\).

The dynamical matrix is constructed from the force constants as
\begin{equation}\label{dynamic_matrix}
    D_{j\alpha,j^\prime\alpha^\prime}(\mathbf{q}) = \frac{1}{\sqrt{m_j m_{j^\prime}}} \Phi_{j\alpha,j^\prime\alpha^\prime} \, e^{i\mathbf{q}\cdot(\mathbf{R}_{j^\prime}-\mathbf{R}_j)},
\end{equation}
where \(m_j\) and \(m_{j^\prime}\) are the atomic masses and \(\mathbf{q}\) is the phonon wave vector.

By diagonalizing the dynamical matrix, the phonon frequencies \(\omega_{\mathbf{q}k}\) and eigenvectors \(\mu_{j\alpha k}(\mathbf{q})\) are obtained:
\begin{equation}\label{diagonalization}
    \sum_{j^\prime \alpha^\prime} D_{j\alpha,j^\prime\alpha^\prime}(\mathbf{q}) \, \mu_{j^\prime \alpha^\prime k}(\mathbf{q}) = \omega_{k}^2(\mathbf{q}) \, \mu_{j\alpha k}(\mathbf{q}).
\end{equation}

\section{Comparison with DFT}

\subsection{Phonon frequency and Huang-Rhys factor}

\begin{figure*}[htbp]
	\centering
	\includegraphics[width=1\textwidth]{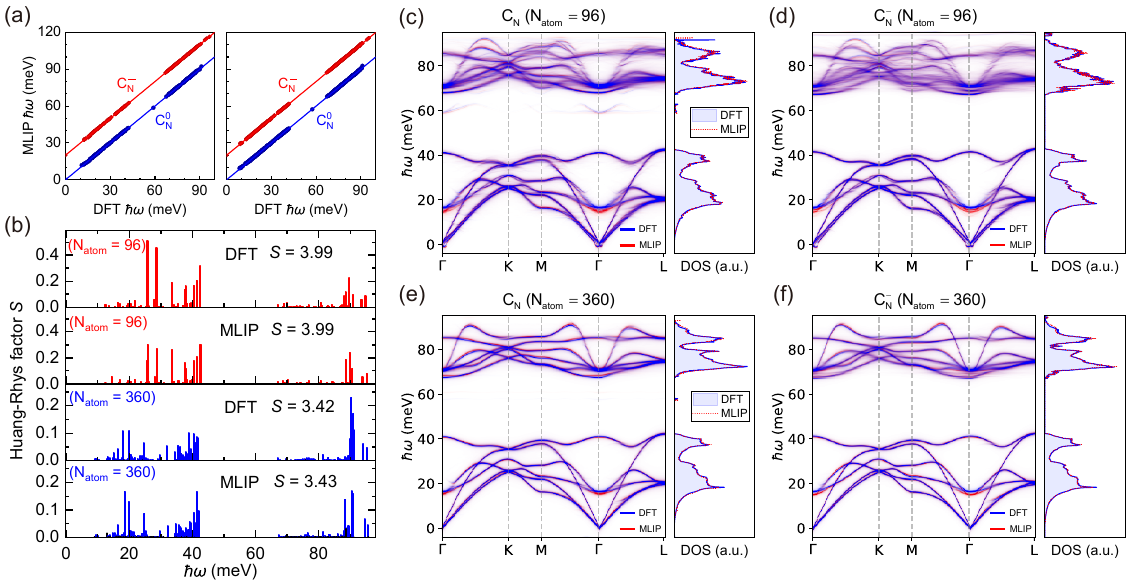}
            \caption{Benchmarking the phonon prediction accuracy of the MLIP against DFT for neutral ($\text{C}_\text{N}^0$) and negatively charged ($\text{C}_\text{N}^{-}$) defects in 96-atom and 360-atom supercells.  
            (a) Comparison of phonon frequencies between MLIP and DFT. The frequencies of the $\text{C}_\text{N}^{-}$ defect predicted by MLIP are shifted upward by 20~meV along the vertical axis.  
            (b) Comparison of the Huang-Rhys factors $S_k$.  
            (c)-(f) Unfolded phonon band structures and densities of states (DOS) projected onto the primitive cell (4 atoms). In the phonon band structures, red lines denote MLIP predictions, while blue lines represent DFT results. In the DOS plots, red dashed lines correspond to MLIP, and blue solid lines to DFT.
            }
	\label{FIG2}
\end{figure*}

First-principle calculations of point defect energies and phonon properties are typically performed using supercells. In such cases, many phonon eigenvalues from the Brillouin zone are folded onto the $\Gamma$ point (\( \mathbf{q} = 0 \)), allowing the defect-induced vibrational properties to be analyzed using only the phonon modes at $\Gamma$. For the $\text{C}_{\text{N}}$ defect, we trained several MLIPs models using 96-atom and 360-atom supercells with neutral and -1 charged states, respectively. Each of them with a training set composed of 40 supercells of the same size. The root-mean-square error (RMSE) of the atomic forces in the training set is $3.9~\text{meV/\AA}$, while the RMSE in the validation set is $5.7~\text{meV/\AA}$. For comparison, we also performed DFT-based phonon calculations for both supercell sizes. The resulting phonon frequencies are shown in Fig.~\ref{FIG2}(a), where the horizontal and vertical axes represent the DFT and MLIP-calculated frequencies, respectively. Blue points denote the $\Gamma$-point phonon frequencies (expressed as $\hbar\omega$) for the neutral defect $\text{C}_\text{N}^0$, while red points correspond to the singly charged defect $\text{C}_\text{N}^{-}$. To distinguish the two charge states visually, the frequencies of $\text{C}_\text{N}^{-}$ are shifted upward by 20~meV. The red and blue dashed lines show perfect agreement.  

In particular, for the neutral defect $\text{C}_\text{N}^0$, the localized defect mode appears around 60~meV, whereas for $\text{C}_\text{N}^{-}$, the localized mode is near 95~meV. The excellent agreement between MLIP and DFT results for both 96-atom and 360-atom supercells confirms that the MLIP accurately captures the phonon behavior near the defect, with only 40 training sets including 40 defect configurations with different bond lengths.

However, it should be noted that since frequencies at $\Gamma$ are sorted in ascending order after diagonalization, the apparent alignment of frequencies in Fig.~\ref{FIG2}(a) may not imply perfect agreement between MLIP and DFT results. Mode reordering may still occur. For instance, if the MLIP incorrectly predicts a phonon mode, it may result in the frequency being overestimated and coincidentally aligned with another mode in the DFT calculation. 

To assess the impact of such mode mismatches and further verify the practical reliability of the MLIP, we introduce the Huang-Rhys factor \(S\), which is defined as
\begin{equation}\label{HR_factor}
    S_k = \frac{1}{2\hbar} \omega_k \Delta Q_k^2 ,
\end{equation}
where \(\omega_k\) is the phonon frequency of the \(k^\text{th}\) mode, and \(\Delta Q_k\) represents the lattice relaxation along the normal coordinate of that mode from the initial state to the final state. It is expressed as
\begin{equation}\label{delta_Q}
    \Delta Q_k = \sum_{j, \alpha } \mu_{j\alpha k} \sqrt{m_j} \Delta R_{j\alpha},
\end{equation}
where \(\mu_{j\alpha k}\) is the phonon eigenvector component for atom \(j\) in Cartesian direction \(\alpha\), \(m_j\) is the atomic mass, and \(\Delta R_{j\alpha} = R^f_{j\alpha} - R^i_{j\alpha}\) is the atomic displacement from the initial to final structure.

For the GaN system, the distribution of \(\Delta Q_k\) spans both low and high frequency regions, as shown in Fig.~\ref{FIG2}(b). The red bars correspond to the Huang-Rhys factors computed from the 96-atom supercell, while the blue bars are from the 360-atom supercell. The total Huang-Rhys factor \(S = \sum_k S_k\) is also labeled in the figure. According to Fig.~\ref{FIG2}(b), most Huang-Rhys factors obtained from MLIP closely match the DFT values for the same supercell size, though minor discrepancies are observed. These differences can be categorized into two cases. 

The first case is that some modes have similar frequencies but differ in their Huang-Rhys factors. For example, in the 96-atom supercell, a low-frequency mode at \(\hbar\omega = 25.8\)~meV yields \(S = 0.51\) from DFT, whereas the MLIP predicts \(S = 0.30\). This discrepancy arises from the near-degeneracy of phonon modes with similar frequencies and differing vibrational patterns. Such mode mixing is common in phonon calculations, where minor variations in constructing the dynamical matrix can lead to different orientations of nearly degenerate modes. Fortunately, this does not significantly affect the calculation of multiphonon transitions.

The second case is that some modes yield similar Huang-Rhys factors but different frequencies. For example, in the high-frequency region of the 360-atom supercell, a mode with \(\omega = 90.2\)~meV in DFT corresponds to \(\omega = 88.3\)~meV in the MLIP prediction. The small difference in frequencies indicates a small deviation in the predicted interatomic forces, yet still within acceptable accuracy for practical applications.

\subsection{Band unfolding}

According to Eqs.~(\ref{HR_factor}) and~(\ref{delta_Q}), the Huang-Rhys factor depends on the lattice relaxation \(\Delta R\) from the initial to the final structure, as well as its projection \(\Delta Q_k\) onto each vibrational normal mode. Since this analysis is based on a specific structural change pathway, it does not ensure that the MLIP accurately describes other vibrational directions that are orthogonal to \(\Delta Q_k\) in the full vibrational space. For example, \(\Delta Q\) is projected along each vibrational mode, and for some modes \(\Delta Q_k\) are nearly zero, leading to negligible contributions to the Huang-Rhys factor. As a result, the MLIP accuracy along these orthogonal directions remains unverified.

To assess the overall accuracy of the MLIP in a more comprehensive manner, we performed phonon band unfolding using the \texttt{UPHO} package \cite{upho}. Phonon dispersions calculated from both the 96-atom and 360-atom supercells were unfolded back to the 4-atom primitive cell. The unfolded phonon band structures computed using DFT and MLIP are shown in Figs.~\ref{FIG2}(c)--\ref{FIG2}(f), where blue lines represent DFT results and red lines correspond to MLIP predictions. For the \(\text{C}_\text{N}\) system, it shows excellent agreement across high-symmetry paths, with only minor deviations in the lowest optical branch at the \(\Gamma\) point. This indicates that the MLIP provides a highly accurate global description of phonon properties.

Since defect-induced localized vibrational mode contributes only a small fraction to the total phonon states, their spectral weight becomes nearly invisible after band unfolding. To further validate the MLIP’s description of the defect mode, we also compared the phonon density of states (DOS) obtained from DFT and MLIP. As shown in Figs.~\ref{FIG2}(c)--\ref{FIG2}(f), the blue solid lines represent the DFT results, and the red dashed lines indicate MLIP predictions. For the \(\text{C}_\text{N}\) defect, the localized mode appears around 60~meV in the neutral state and shifts to approximately 94~meV in the negatively charged state, with the defect-induced DOS peaks from DFT and MLIP in excellent agreement.

For the \(\text{Li}_{\text{Zn}}\) system, the unfolded phonon bands and DOS are provided in the Supplemental Material. Similar to the \(\text{C}_\text{N}\) case, it shows excellent agreement between the MLIP and DFT results, further confirming the generality of the approach.

\section{Application}
\subsection{Radiative luminescence lineshape}
Taking the \(\text{C}_{\text{N}}\) defect as an example, we consider a radiative transition process in which an electron in the conduction band minimum (CBM) is captured onto the defect level and emits a photon. Our goal here is to demonstrate the applicability of MLIP to multiphonon optical processes rather than predicting optical spectra that perfectly match the experiment, therefore, the defect geometries in the two charge states involved in this calculation are obtained using the PBE functional. Since PBE usually underestimates lattice relaxations compared to hybrid functional, the resulting photoluminescence (PL) spectrum may differ from a full hybrid-functional-based result.

Under the Condon approximation \cite{Alkauskas_2014_NV}, the PL intensity \(I(\hbar\omega)\) at \(T = 0\)~K is given by
\begin{equation}\label{PL_intensity}
    I(\hbar\omega) \propto \omega^3 \sum_{n} \left| \left< \chi_{i0} \middle| \chi_{fn} \right> \right|^2  \delta (E_\text{ZPL} - E_{fn} - \hbar\omega),
\end{equation}
where \(\hbar\omega\) is the emitted photon energy, \(E_\text{ZPL}\) is the zero-phonon line energy, \(i\) and \(f\) denote the initial and final electronic states, and \(n\) is the number of phonons emitted in the final state. \(\chi_{i0}\) is the vibrational ground-state wavefunction of the initial configuration. 

Direct evaluation of Eq.~(\ref{PL_intensity}) is typically impractical, and the PL spectrum is instead computed using the Fourier transform of a generating function. We write the intensity as \(I(\hbar\omega) = C\omega^3 A(\hbar\omega)\), where \(C\) is a normalization constant and \(A(\hbar\omega)\) is the spectral function, defined as
\begin{equation}\label{spectral_function}
    A(E_\text{ZPL} - \hbar\omega) = \frac{1}{2\pi} \int_{-\infty}^{+\infty} G(t) \, e^{i\omega t - \gamma |t|} dt,
\end{equation}
where \(\gamma\) is a broadening parameter used to match experimental linewidths. The generating function \(G(t)\) is defined as
\begin{equation}\label{gf}
    G(t) = e^{S(t) - S(0)},
\end{equation}
with
\begin{subequations}
    \begin{align}
        S(t) &= \int_{0}^{+\infty} S(\hbar\omega) \, e^{-i\omega t} \, d(\hbar\omega), \\
        S(0) &= \int_{0}^{+\infty} S(\hbar\omega) \, d(\hbar\omega) = \sum_k S_k.
    \end{align}
\end{subequations}

Different from the discrete Huang-Rhys factors \(S_k\), \(S(\hbar\omega)\) is a continuous spectral distribution fundamental to the PL process, defined as
\begin{equation}\label{S_hw}
    S(\hbar\omega) = \sum_k S_k \, \delta (\hbar\omega - \hbar\omega_k),
\end{equation}
where the $\delta$ function is typically broadened using a Gaussian function in numerical method.

The spectral function \(S(\hbar\omega)\) is highly sensitive to the distribution of Huang-Rhys factors. Therefore, achieving a converged \(S(\hbar\omega)\) requires phonon calculations on large supercells. However, performing such calculations using conventional finite-displacement methods becomes prohibitively expensive for supercells containing more than 1000 atoms. For supercell including defect, there are techniques that can significantly accelerate phonon calculations in large supercells, such as the force-constant embedding scheme proposed by Alkauskas \cite{Alkauskas_2014_NV,Alkauskas_2021}.

In the finite-displacement method of phonon calculations, the force response on all atoms is obtained by displacing one atom in a specific direction and performing a single electron self-consistent field (SCF) calculation. The embedding approach relies on two assumptions. First, if \(\Psi'\) and \(\Psi\) denote the force-constant tensors of a defect and a bulk supercell in the same size, respectively, then \(\Delta\Psi = \Psi' - \Psi\) is assumed to be spatially localized, with the defect-induced perturbation decaying exponentially with distance. Second, the force response induced by displacing a single atom also decays exponentially with distance. Based on these two assumptions, the force constants of a large supercell can be constructed by embedding data from two smaller supercells: one containing the defect and one bulk. For instance, the force constants of a \(\sim 10^5\)-atom \(\text{C}_\text{N}\) supercell can be built using only the 360-atom defect supercell and the corresponding 360-atom bulk supercell, as the force responses of a displaced atom within the 360-atom size can be evaluated from the forces in these two supercells, while those beyond the size are supposed to shrink to zero.

However, special care must be taken when embedding the defect force constants into the large supercell, due to the periodic boundary condition of the limited-size defect supercell. Atoms near the boundary may experience artificial interactions due to periodic images, which do not exist in the larger supercell. To address this, a cut-off radius \(r_d\) must be defined to discard force constants near the defect supercell boundary. Typically, \(r_d\) should be less than one quarter of the defect supercell size. Consequently, the limited-size defect supercell should also be sufficiently large, which limits the applicability of this method for more time-consuming DFT calculations, such as hybrid functional.

\begin{figure*}[htbp]
	\centering
	\includegraphics[width=1.0\textwidth]{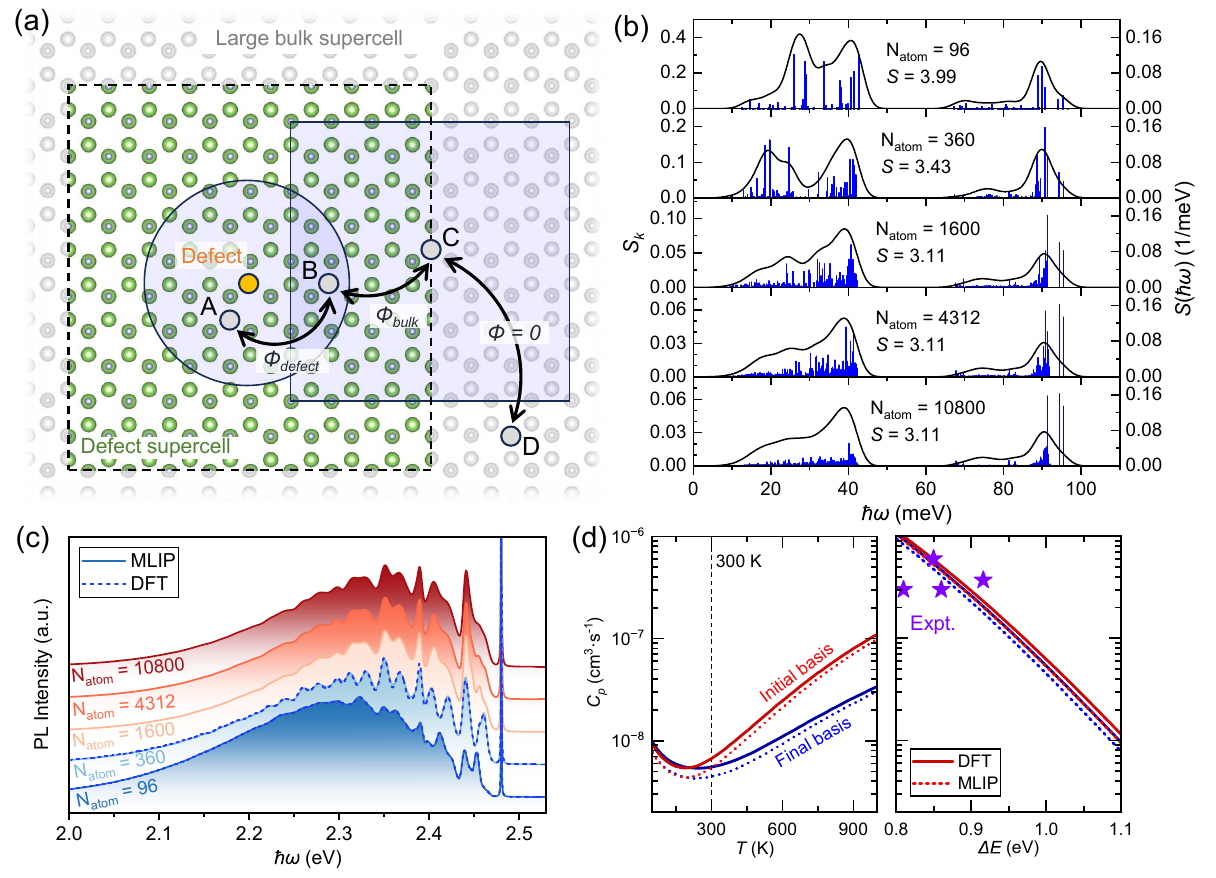}
	\caption{
        Application of MLIP-accelerated phonon calculations.  
        (a) Schematic of constructing force constants in a large supercell using MLIP. The yellow circle represents the defect site, the transparent blue circular region indicates the cut-off radius \(r_d\) around the defect supercell, and the transparent blue square marks the boundary \(r_b\) beyond which MLIP predictions are truncated to zero.  
        (b) Convergence test of the Huang-Rhys factors \(S_k\) and spectral function \(S(\hbar\omega)\) of the \(\text{C}_{\text{N}}\) defect as a function of supercell size.  
        (c) Radiative PL lineshapes calculated for different supercell sizes; blue dashed lines indicate DFT results.  
        In (b) and (c), the 96-atom and 360-atom supercells are computed using MLIPs trained on cells of the same size, while larger supercells are constructed using the embedding approach shown in (a).  
        (d) Nonradiative capture coefficient \(C_p\) of the \(\text{C}_{\text{N}}\) defect as a function of temperature \(T\) and transition energy \(\Delta E\). Dashed lines represent MLIP predictions, solid lines correspond to DFT calculations, red lines are computed using the initial-state basis, blue lines with the final-state basis, and the purple star denotes the experimental value.
        }
	\label{FIG3}
\end{figure*}

In contrast, MLIPs provide a more scalable alternative. As shown in the previous section, phonon calculations in supercells containing hundreds of atoms can be performed using only 40 training sets. Based on this advantage, we propose an improved embedding scheme using MLIPs. The steps are as follows:  
(i) train an MLIP using 40 defect supercells;  
(ii) compute the force constants of the defect supercell using the trained MLIP;  
(iii) compute the force constants of a large bulk supercell using the same MLIP;  
(iv) embed the defect force constants within radius \(r_d\) into the large supercell.

A schematic illustration of this process is shown in Fig.~\ref{FIG3}(a). The force constants of the large bulk supercell, \(\Phi_{\text{bulk}}\), are directly computed using the MLIP for defect. Since MLIP predictions are constrained by the size of supercell used in the training set, forces beyond a boundary radius \(r_b\) are predicted as zero, which automatically satisfies the second assumption made in the embedded scheme. The atomic structure of the defect supercell is then embedded into the large supercell, and within the cut-off radius \(r_d\), the bulk force constants \(\Phi_{\text{bulk}}\) are replaced by the defect force constants \(\Phi_{\text{defect}}\). Figure~\ref{FIG3}(a) also illustrates the source of each interatomic force constant: if both atoms lie within \(r_d\), such as A and B, then \(\Phi_{AB} = \Phi_{\text{defect}}\); if either atom lies outside \(r_d\), \textit{e.g.}, atoms B and C, then \(\Phi_{BC} = \Phi_{\text{bulk}}\); and if the interatomic distance exceeds \(r_b\) (\textit{i.e.}, outside the solid-line box in Fig.~\ref{FIG3}(a)), the force constant is set to zero.

Compared to conventional embedding approaches, this method offers two advantages. First, \(\Phi_{\text{bulk}}\) for the large supercell can be obtained directly from MLIP without performing extra DFT calculations. Second, since the defect supercell inherently contains many bulk-like atomic configurations, the force constants in the bulk large supercell also maintain high accuracy. Additional benchmarking results are provided in the Supplemental Material.

The dynamical matrix can be constructed from the force constants through Eq.~(\ref{dynamic_matrix}). Since we are interested in vibrational modes at the \(\Gamma\) point (\( \mathbf{q} = 0 \)) in large supercells, the Fourier transform is unnecessary. However, for supercells containing more than \(10^5\) atoms, direct diagonalization of the dynamical matrix becomes computationally challenging. Since most off-diagonal elements in such large dynamical matrices are zero, making them highly sparse. We therefore employ the \texttt{SLEPc} library \cite{slepc} to efficiently diagonalize the sparse matrix.

We continue to use the \(\text{C}_{\text{N}}\) defect calculated with the PBE functional as a benchmark. Figure~\ref{FIG3}(b) shows the convergence behavior of the Huang-Rhys factors (left vertical axis) in blue bars and the spectral function \(S(\hbar\omega)\) (right vertical axis) in black lines with increasing supercell size. For the 96-atom and 360-atom supercells, phonon frequencies and eigenvectors are directly obtained using MLIPs trained on the respective supercell size. For the 1600-atom, 4312-atom, and 10800-atom supercells, the phonon properties are obtained using our force-constant embedding approach, with the MLIP trained on the 360-atom supercell.

Two localized phonon modes associated with the \(\text{C}_{\text{N}}\) defect appear around \(94\,\text{meV}\) in Fig.~\ref{FIG3}(b). Due to the embeded defect structure within the radius are the same (360-atom), the corresponding Huang--Rhys factors remain nearly unchanged. The shape of the spectral function \(S(\hbar\omega)\) directly determines the PL lineshape through Eq.~(\ref{spectral_function}). From Fig.~\ref{FIG3}(b), we observe that \(S(\hbar\omega)\) becomes converged for the 1600-atom supercell.

The resulting PL spectra are shown in Fig.~\ref{FIG3}(c). The dashed lines correspond to results obtained from DFT, while the solid lines are computed using MLIP-predicted phonon properties. All PL spectra are calculated with the broadening factor of \(\gamma = 0.3\,\text{meV}\) in Eq.~(\ref{spectral_function}) and the delta function with a Gaussian width of \(2\,\text{meV}\) in Eq.~(\ref{S_hw}). In line with the convergence of \(S(\hbar\omega)\) at the 1600-atom supercell in Fig.~\ref{FIG3}(b), the PL spectra are also found to converge at that size.

\subsection{Nonradiative multiphonon transitions}
The rate of nonradiative multiphonon transitions can be described using Fermi’s golden rule \cite{huang1981lattice,passler1974description}:
\begin{equation}\label{fermi_golden_rule}
    r = \frac{2\pi}{\hbar} \rho(E_{im}) \left| \left< \Phi_{im} \left| H_{eL} \right| \Phi_{fn} \right> \right|^2 \delta(\Delta E + E_m - E_n),
\end{equation}
where \( H_{eL} \) denotes the electron–phonon interaction Hamiltonian, and \( \Phi \) is the total wavefunction of the lattice system, including both the electronic wavefunction \( \varphi \) and the phonon wavefunction \( \chi \). The indices \( i \) and \( f \) label the initial and final electronic states, while \( m \) and \( n \) are the number of phonons in the initial and final states, respectively. The phonon energy of the initial state is given by \( E_{im} = \hbar \omega_i (m + 1/2) \), and \( \Delta E \) is the transition energy from initial state to final state. The occupation of phonons in the initial state, \( \rho(E_{im}) \), follows the Boltzmann distribution.

Under the static coupling approximation \cite{passler1974description,passler1975description} and linear coupling approximation \cite{huang1981lattice} of the electron–phonon interaction, the matrix element becomes:
\begin{equation}\label{separation}
    \left< \Phi_{im} \left| H_{eL} \right| \Phi_{fn} \right> = \sum_k W_k \int dQ \ \chi_m^*(\bar{Q}) Q_k \chi_n(\bar{\bar{Q}}),
\end{equation}
where \( W_k = \left< \varphi_i \left| \partial H_{eL}/\partial Q_k \right| \varphi_f \right> \) is the electron–phonon coupling matrix element, and \( Q \) represents the normal mode coordinate. \( \bar{Q} \) is the initial state normal coordinate and \( \bar{\bar{Q}} \) is the final state normal coordinate.

In our previous work \cite{zhou2025}, we considered the defect phonon renormalization during multiphonon transitions involving both initial and final states. The relation between two sets of phonons is described by the Duschinsky rotation \cite{duschinsky1937importance,baiardi2013general}:
\begin{equation}\label{Duschinsky}
    J_{kk^\prime} = \sum_{j, \alpha} \bar{\mu}_{j\alpha k} \, \bar{\bar{\mu}}_{j\alpha k^\prime},
\end{equation}
where \( \bar{\mu}_{j\alpha k} \) is the eigenvector for the initial states and \( \bar{\bar{\mu}}_{j\alpha k^\prime} \) is for final states. According to Eq.~(\ref{Duschinsky}), nonradiative transition rate requires phonon calculations for both initial and final states, which results in significant computational expense. As a result, various approximations have been introduced to balance accuracy and efficiency \cite{alkauskas2014first,shi2012,shi2015,turiansky2021nonrad,turiansky2025}. With MLIP, we are now able to substantially reduce the computational cost of phonon calculations with only 40 training sets.

To be consistent with our previous work \cite{zhou2025}, here we used hybrid-functional DFT calculations \cite{heyd2003hybrid,10.1063/1.2204597} on the same 96-atom supercell containing the $\text{C}_{\text{N}}$ defect. Following the previous training procedure, we constructed MLIPs for both the neutral defect $\text{C}_{\text{N}}^0$ and the negatively charged defect $\text{C}_{\text{N}}^-$, each using a training set of 40 perturbed structures with a maximum displacement radius of $r_{\mathrm{max}} = 0.04~\text{\AA}$.For $\text{C}_{\text{N}}^0$, the RMSE for the atomic forces was $3.4~\text{meV/\AA}$ on the training set and $6.6~\text{meV/\AA}$ on the validation set. For the negatively charged $\text{C}_{\text{N}}^-$ defect, the force RMSE was $3.1~\text{meV/\AA}$ on the training set and $4.2~\text{meV/\AA}$ on the validation set.

We followed the methodology of our previous work \cite{zhou2025} in the nonradiative capture coefficient calculations. We considered the process of a hole at the valence band maximum (VBM) captured by a defect, with the transition energy of $\Delta E = 1.13~\text{eV}$. Using the final state as the vibrational basis, the MLIP-accelerated capture coefficient was found to be $C_p = 4.45 \times 10^{-9}~\text{cm}^3\text{ s}^{-1}$ at $T = 300~\text{K}$, close to the DFT result of $C_p = 5.52 \times 10^{-9}~\text{cm}^3\text{ s}^{-1}$. When using the initial state as the vibrational basis, the MLIP result was $C_p = 5.45 \times 10^{-9}~\text{cm}^3\text{ s}^{-1}$, while the DFT prediction was $C_p = 7.04 \times 10^{-9}~\text{cm}^3\text{ s}^{-1}$. In both cases, the MLIP results agree with the DFT results within a factor of 1.3.

The dependence of the capture coefficient on both temperature $T$ and transition energy $\Delta E$ is shown in Fig.~\ref{FIG3}(d). In this figure, red lines are results obtained using the initial state basis, and blue lines correspond to the final state basis. Solid lines represent DFT calculations, while dashed lines indicate MLIP predictions. The MLIP and DFT results reach a close agreement across a wide range of temperature and transition energy. Experimental data points from the literature are also included in Fig.~\ref{FIG3}(d) as purple stars, indicating that the MLIP-accelerated rates can achieve near-DFT accuracy.

\subsection{Time consumption}

\begin{figure}[h]
	\centering
	\includegraphics[width=0.45\textwidth]{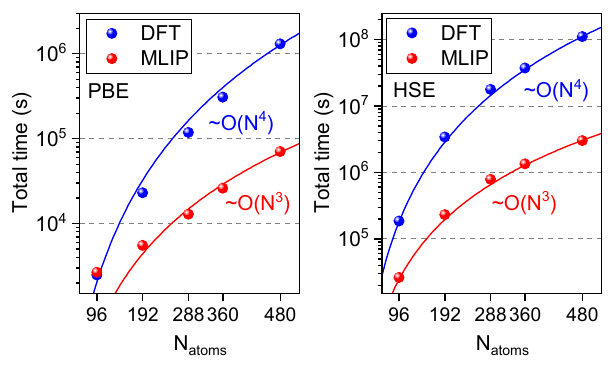}
	\caption{Comparison of the total computational time required for a full phonon calculation using either DFT or MLIP across different supercell sizes. Blue solid lines represent quartic fits of the form $AN^4$, and red solid lines correspond to cubic fits of the form $BN^3$, with $A$ and $B$ as fitting parameters. The left panel shows results using the PBE functional, while the right panel uses the HSE functional.}
	\label{FIG4}
\end{figure}

Historically, the high computational cost has posed a significant barrier to phonon calculations in large supercells. In DFT, the SCF calculation time $t$ scales approximately as $t \propto N^3$, where $N$ is the number of atoms in the supercell \cite{fonsecaguerraOrderNDFTMethod1998}. For the finite displacement method, the low symmetry of defect structures typically requires $\sim 6N$ perturbed structures. Consequently, the total computational cost scales as $N^4$.

As demonstrated in previous sections, our MLIP approach achieves excellent accuracy in phonon-related quantities, including phonon band structures, vibrational frequencies, nonradiative transition rates, and PL spectra. Importantly, for supercells considered in this work, only 40 training are found to be sufficient. Therefore, the training data preparation (based on DFT) scales as $N^3$, representing a reduction in computational cost by a factor of $N$ compared to a full DFT-based phonon calculation.

To evaluate the computational efficiency quantitatively, we benchmarked the wall time for phonon calculations using both DFT and MLIP across supercells of increasing size. The tests were performed on an Intel(R) Xeon(R) Platinum 8375C CPU (64 cores) and an NVIDIA GeForce RTX 4090 D GPU. The results for the $\text{C}_{\text{N}}$ defect system are shown in Fig.~\ref{FIG4}, where Fig.~\ref{FIG4}(a) corresponds to calculations with the PBE functional, and Fig.~\ref{FIG4}(b) uses the HSE functional. The vertical axis in Fig.~\ref{FIG4} represents the total time cost. For DFT, this corresponds to the total time of approximately $3N$ SCF calculations. For MLIP, the total time consists of (i) 40 SCF calculations for training data generation and (ii) the MLIP training time on GPU (converted into equivalent CPU time for comparison). The prediction time for MLIP is negligible (on the order of one minute) and thus omitted. In both panels, the blue and red curves represent fitted scaling trends: $AN^4$ for DFT and $BN^3$ for MLIP, where $A$ and $B$ are fitting parameters. Most data points closely follow the expected scaling. For the PBE functional, MLIP becomes more than an order of magnitude faster than DFT for supercells larger than 300 atoms. When using the HSE functional, the same level of acceleration is achieved for supercells with more than 100 atoms.

However, for small supercells, DFT remains more time-efficient. In particular, for the 96-atom PBE case in Fig.~\ref{FIG4}(a), the red point (MLIP) lies above the blue point (DFT), indicating that the MLIP training time exceeds the total SCF time of the DFT phonon calculation. This is expected, as PBE is computationally inexpensive and MLIP training cost does not depend on the functional, resulting in a deviation from the ideal $N^3$ trend.

\section{Conclusions}
In conclusion, we demonstrate that high-accuracy phonon predictions for defect-including supercells can be achieved using defect-specific MLIPs trained on a limited dataset of stuctures. Our test examples on the $\text{C}_\text{N}$ in GaN and $\text{Li}_\text{Zn}$ in ZnO show that MLIP-based phonon calculations are more than an order of magnitude faster than DFT using finite displacement method. This method can be easily integrated into existing DFT workflows, such as the embedding scheme for 10800-atom supercell phonon calculations, and largely improve the computational efficiency of multiphonon transitions meanwhile maintaining the DFT-level accuracy. We anticipate that this method will be promising in studies with defect phonons and offer a practical solution for high-throughput predictions of phonon properties in defect systems.

\section{Acknowledgments}
This work was supported by National Natural Science Foundation of China (12334005, 12174060, 12188101 and 12404089), National Key Research and Development Program of China (2022YFA1402904 and 2024YFB4205002), Science and Technology Commission of Shanghai Municipality (24JD1400600 and Explorer project 24TS1400500), and Project of MOE Innovation Platform.


\bibliography{apssamp}

\end{document}